\documentclass{appolb}

\usepackage{graphicx}

\begin{document}

\title{ Paradox of integration - mean field approach
\thanks{Thanks
}
}

\author{
Krzysztof~Ku{\l}akowski, Piotr Gronek
\address{AGH University of Science and Technology, Faculty of Physics and Applied Computer Science, al.~Mickiewicza 30, 30-059 Krak\'ow, Poland}
\\[.5cm]
Alfio Borz\`i
\address{Institut f\"ur Mathematik, Universit\"at W\"urzburg, Emil-Fischer-Strasse 30, 97074~W\"urzburg, Germany}
\\[.5cm]
}

\maketitle
\begin{abstract}
Recently a computational model has been proposed of the social integration, as described in sociological terms by Peter Blau. In this model, actors praise or critique each other, and these actions influence their social status and raise negative or positive emotions. The role of a self-deprecating strategy of actors with high social status has also been discussed there. Here we develop a mean field approach, where the active and passive roles (praising and being praised, etc.) are decoupled. The phase transition from friendly to hostile emotions has been reproduced, similarly to the previously applied purely computational approach. For both phases, we investigate the time dependence of the distribution of social status. There we observe a diffusive spread, which - after some transient time - appears to be limited from below or from above, depending on the phase. As a consequence, the mean status flows.

\end{abstract}



\section{Introduction}
\label{S1}

Theory of social integration has been provided by Peter Blau \cite{blau}. During the integration process, the group members simultaneously compete for social status and play the role of the audience. The paradoxical aspect is that those who gain high social status are not accepted because others are afraid of being dominated. As was described by Blau, people with high status are more inclined to bother with acceptance of others; this is achieved by an application of a self-deprecating strategy. Slightly reducing their overwhelming advantage in status, for example by self-mockery, they gain sympathy of the audience. \\

Recently, a computational version of the Blau theory has been proposed \cite{my}. In this model, actors praise or critique each other, and these actions influence their social status and raise negative or positive emotions. Namely, if Z praises Y, then {\it (i)} Z is more liked, {\it (ii)} the status of Y increases, {\it (iii)} the status of Z decreases. Conversely, if Z criticizes Y, then {\it (i)} Z is less liked, {\it (ii)} the status of Y decreases, {\it (iii)} the status of Z increases. Two results of \cite{my} are to be highlighted here. First, there is a sharp transition between the phase where most actors are liked and the phase where most are disliked. Second, if the probability of praising by Z increases with the social status of Z, the transition becomes smooth. Such an increase is just the self-deprecating strategy. All these results have been obtained with the Monte-Carlo method, as ordered pairs of actors have been selected randomly. Once selected, the decision to praise or to criticize has been taken by a calculation of a model work fuction. \\

The aim of this work is to develop a deterministic mean-field approach of the problem. Here the work function is averaged over all actors. Further, the actor state, being praised or criticized, is also decided by another average of the work function. In this way, "praising or not" and "being praised or not" are two streams of decisions, which are coupled only by the relations of single actors with the whole group. \\

In the next section, the mean field model is described, with atention paid to its comparison with the preceding computational version \cite{my}. Section 3 provides the results; we will show that the model behavior is surprisingly rich. Last section is devoted to discussion and conclusions.

\section{The model}

In the computational formulation \cite{my}, an ordered pair of actors (Z,Y) had been chosen randomly, from a uniform distribution and with replications. The decision if Z had to praise Y or to criticize her/him (below we write on males for brevity) was taken, basing on the work function

\begin{equation}
f(Z,Y)))=-p'(Z)+\frac{1-p'(Z)}{N-1}v(A(Y))
\label{f1}
\end{equation}
where

\begin{equation}
p'(Z)=\frac{2p}{1+\gamma ^{A(Z)}}
\label{f2}
\end{equation}
$p\in (0,1)$ was a parameter which measured the willingness to criticize, $N$ was the number of actors, $v(A)$ was the number of actors with status $A$, and $\gamma$ allowed to distinguish the case with ($\gamma=2$) and without  ($\gamma=1$) the self-deprecating strategy. Hence, the work function $f$ depended on time both through individually assigned status $A$ for each actor (Eq. \ref{f2}), and through the time-dependent distribution $v(A)$ (Eq. \ref{f1}).\\

In this work, instead of random sequential selection of pairs, we work on the distribution of status $v(A,t)$ and on emotions $x(A,t)$. Here, the status variable $A$ is no more assigned to individual actors; it is just an argument of the functions $v$ and $x$, and it does not depend on time. For each value $A$ of the status, two average work functions are calculated to state if an agent of this status is praised or criticized, and if he is praising or criticizing. First issue is decided, basing on the work function 
$f_1(A,t)$, defined as

\begin{equation}
f_1(A,t)=\frac{1}{N}\sum_B v(B,t) f(B,v(A,t)) 
\label{f3}
\end{equation}
and the second issue - basing on $f_2(A)$, defined as

\begin{equation}
f_2(A,t)=\frac{1}{N}\sum_B v(B) f(A,v(B,t))
\label{f4}
\end{equation}
There, the function $f$ is defined as

\begin{equation}
f(A,v(B,t))=-p'(A)+\frac{1-p'(A)}{N-1}v(B,t)
\label{f5}
\end{equation}
and 

\begin{equation}
p'(A)=\frac{2p}{1+\gamma ^{A}}.
\label{f6}
\end{equation}
If $f_1(A,t)>0$, one of agents with status $A$ is praised. Then, the function $v(A)$ is modified as follows: $v(A,t+1)=v(A,t)-1$ and $v(A+1,t+1)=v(A+1,t)+1$. Conversely, if $f_1(A,t)<0$, one of agents with status $A$ is criticized. Then, $v(A,t+1)=v(A,t)-1$ and $v(A-1,t+1)=v(A-1,t)+1$. Further, if $f_2(A,t)>0$, one of agents with status A is praising. Then, $v(A,t+1)=v(A,t)-1$ and $v(A-1,t+1)=v(A-1,t)+1$. Also, $x(A,t+1)=1$. Conversely, if $f_2(A,t)<0$, one of agents with status $A$ is criticizing. Then, $v(A,t+1)=v(A,t)-1$ and $v(A+1,t+1)=v(A+1,t)+1$. Also, $x(A,t+1)=-1$. In one time step, both functions $v(A,t)$ and $x(A,t)$ are updated simultaneously. To keep $v(A,t)$ non negative, the cases when $v(A,t)<2$ are not modified. The rationale of this algorithm is the same as given in the Introduction, and close to the one used in \cite{my}. A slight difference with respect to \cite{my} is introduced with treating the function $x(A,t)$; to shorten the calculations, now it is updated in one step to its limit values $\pm 1$. However, this function plays the role of a slave variable  \cite{masl}, because its evolution is determined by the evolution of $v(A,t)$. Therefore we consider this modification to be marginal. \\

On the contrary to the computational approach in \cite{my}, the mean value of status $<A>=\sum_A Av(A,t)$ is not conserved here. Yet, the system size $\sum_A v(A,t)=N$ remains conserved.\\

\section{The results}

 \begin{figure}[!hptb]
\begin{center}
\includegraphics[width=\columnwidth]{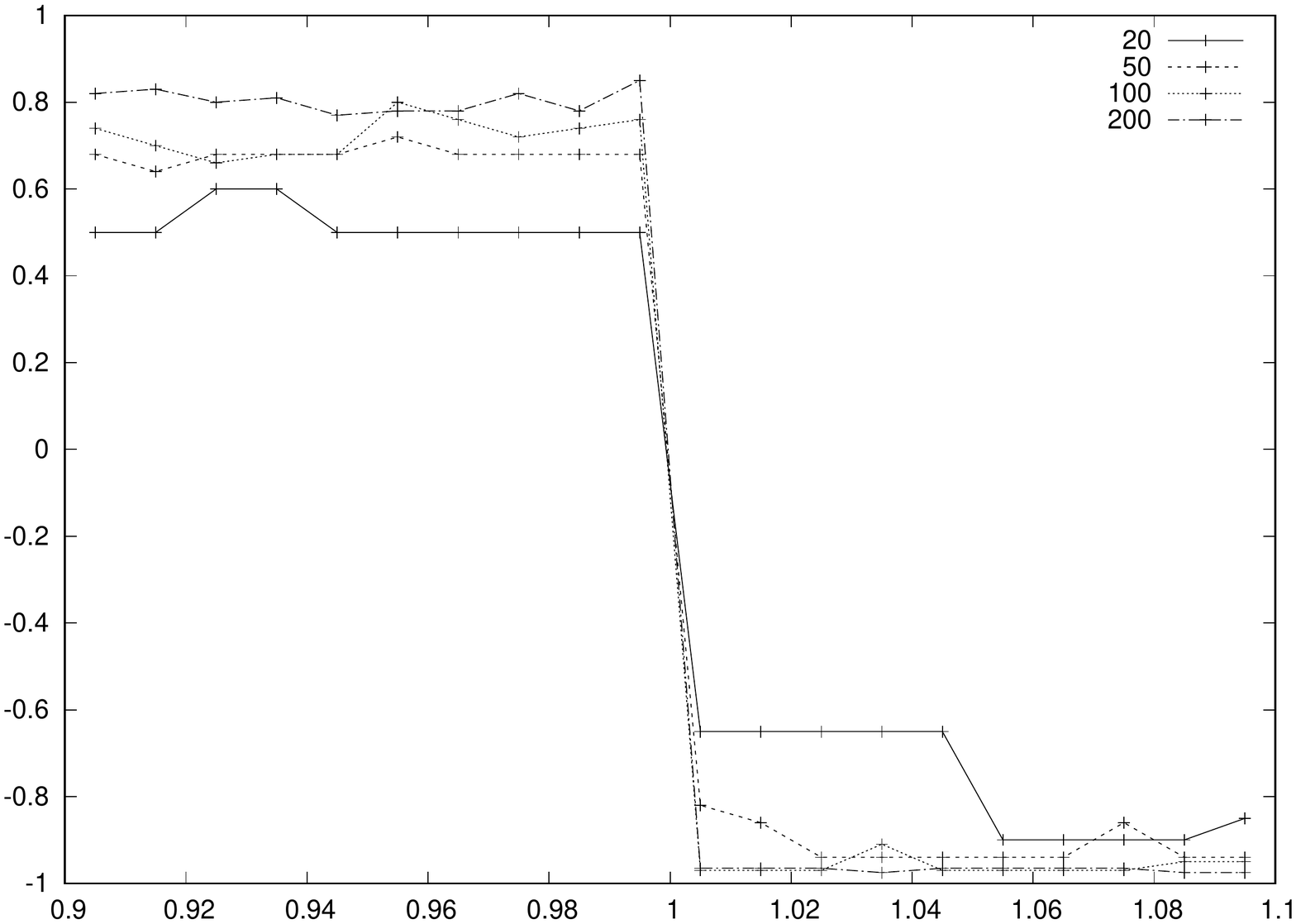}
\caption{The stationary values of $<x>$ against $p(N+2)/2$ for different system sizes $N$ given in the legend.}
\label{fig1}
\end{center}
\end{figure}

As a rule, we use initial conditions $v(A,0)=0$ for all values of $A$ except $A=0$, where $v(0,0)=N$. If $<A> \approx 0$, the initial conditions influence the stationary results only slightly: exemplary test results for $N=100$ are shown in Table 1. \\

\begin{table}[t]
\caption{Results of the test of an influence of the initial status distribution $v(A)$ on the stationary solution $<x(i)>$ for $N=100$, on both sides of the phase transition at $p_c\approx 0.02$. Information is added on the number of time steps.}
\label{t1}
\begin{tabular}{|l|c|c|r|}
\hline
No&initial distribution&$p$=0.01&$p$=0.03\\
\hline
 1&$v(0)$=100&0.76, 3647&-0.93, 1042\\
\hline
2&$v(-1)$=25, $v(0)$=50, $v(1)$=25&0.76, 3622&-0.91, 1550\\
\hline
3&$v(-1)$=50, $v(1)$=50&0.76, 3699&-0.93, 528\\
\hline
4&$v(0)$=50, $v(1)$=50&0.76, 3626&-0.91, 1321\\
\hline
5&$v(-1)$=50, $v(0)$=50&0.80, 3540&-0.93, 1322\\
\hline
\end{tabular}
\end{table}

The transition point $p_c$ is calculated numerically, as dependent on the number of actors $N$. For symmetric initial distributions of status $v(A)$ he results agree well with the formula $p_c=2/(N+1)$ (see Fig. \ref{fig1}). On the other hand, $p_c$ can be derived from the condition on the threshold value of the work function $f=0$:

\begin{equation}
p'=(1-p')\frac{v(A)}{N-1}
\label{f5}
\end{equation}
where $v(A)=2$ is the minimal value of $v$ where the evolution is assumed to be active, and

\begin{equation}
p'=\frac{2p}{1+\gamma ^{<A>}}
\label{f6}
\end{equation}
seems a reasonable approximation. Here we base on a numerical observation, that $<A>$ changes its sign at the transition, together with $<x>$. On the other hand, for asymmetric initial distributions $v(A)$, the critical value $p_c$ is shifted away from $2/(N+1)$ with initial $<A>$ despite the fact that also here both $<A>$ and $<x>$ change their signs at $p=p_c$.\\

When the time evolution of the profile $v(A)$ is observed, two processes can be separated out: a spread around the initial concentration, and a drift. The former is best seen when the initial distribution of status is the Kronecker-like, i.e. $v(A,t=0)=\delta_{A,0}$. Then, the spread is symmetric; the same for positive and negative status $A$. At some time step, however, the symmetry is broken: a limitation of status appears, which blocks its decrease for $p<p_c$ and its increase for $p>p_c$. Examples of both kinds of behavior are shown in Tables 2 and 3. In these examples, the system size is purposefully chosen to be small ($N=9$), to keep a reasonable size of the tables. In both cases, the symmetry is broken already after a few of time steps. However, this is not a rule; for $N=20$, the evolution stops after 96 steps, and the asymmetry appears as late as after 48 steps.\\

Basically, the model behavior depends on four factors: three parameters ($N,\gamma,p$) and the initial distribution of status $v(A,t=0)$. Although here we are concerned mainly with the role of $N$ and $p$, a comment should be added on the remaining two. A decrease of $\gamma$ till 1.1 does not change much; we checked that for $N=50,100$ and the Kronecker-like initial distribution of status, the transition point $p_c$ is not changed. However, for $\gamma=1.0$ (no self-deprecating strategy) and $p<p_c$, the system evolution is captured by a periodic cycle, and the stationary value of $<x>$ cannot be determined. Such cycles do happen occassionally also for $\gamma >1$, and their presence depends on $N,\gamma,p$ and on the initial distribution of $v(A)$. Most often, the observed period length is 4. Cycles have been found also for different initial distributions $v(A).$\\


Table 2. The time evolution, from top to down, of the function $v(A,t)$ for $N$=9 and $p$=0.15 ($p_c=0.2$). In the central column, initially only $v(0)$ is different from zero. The evolution lasts 31 time steps, then stops. The left-right symmetry is broken at 6-th step.\\

\noindent
  0  0  0  0  0  0  0  9  0  0  0  0  0  0  0\\
  0  0  0  0  0  0  1  7  1  0  0  0  0  0  0\\
  0  0  0  0  0  0  2  5  2  0  0  0  0  0  0\\
  0  0  0  0  0  1  1  5  1  1  0  0  0  0  0\\
  0  0  0  0  0  1  2  3  2  1  0  0  0  0  0\\
  0  0  0  0  0  2  1  3  1  2  0  0  0  0  0\\
  0  0  0  0  0  0  4  1  3  0  1  0  0  0  0\\
  0  0  0  0  0  1  2  3  1  1  1  0  0  0  0\\
  0  0  0  0  0  1  1  3  2  1  1  0  0  0  0\\
  0  0  0  0  0  1  2  2  1  2  1  0  0  0  0\\
  0  0  0  0  0  1  1  2  3  0  2  0  0  0  0\\
  0  0  0  0  0  1  2  1  2  2  0  1  0  0  0\\
  0  0  0  0  0  1  0  4  1  1  1  1  0  0  0\\
  0  0  0  0  0  1  1  2  2  1  1  1  0  0  0\\
  0  0  0  0  0  1  2  1  1  2  1  1  0  0  0\\
  0  0  0  0  0  1  0  3  2  0  2  1  0  0  0\\
  0  0  0  0  0  1  1  2  1  2  0  2  0  0  0\\
  0  0  0  0  0  1  2  0  3  0  2  0  1  0  0\\
  0  0  0  0  0  2  0  2  1  2  0  1  1  0  0\\
  0  0  0  0  0  0  3  0  3  0  1  1  1  0  0\\
  0  0  0  0  0  1  1  2  1  1  1  1  1  0  0\\
  0  0  0  0  0  1  1  0  3  1  1  1  1  0  0\\
  0  0  0  0  0  1  1  1  1  2  1  1  1  0  0\\
  0  0  0  0  0  1  1  1  2  0  2  1  1  0  0\\
  0  0  0  0  0  1  1  2  0  2  0  2  1  0  0\\
  0  0  0  0  0  1  2  0  2  0  2  0  2  0  0\\
  0  0  0  0  0  1  0  3  0  2  0  2  0  1  0\\
  0  0  0  0  0  1  1  1  2  0  2  0  1  1  0\\
  0  0  0  0  0  1  1  2  0  2  0  1  1  1  0\\
  0  0  0  0  0  1  2  0  2  0  1  1  1  1  0\\
  0  0  0  0  0  1  0  3  0  1  1  1  1  1  0\\
  0  0  0  0  0  1  1  1  1  1  1  1  1  1  0\\


Table 3. The time evolution, from top to down, of the function $v(A,t)$ for $N$=9 and $p$=0.25 ($p_c=0.2$). The evolution lasts 38 time steps, then it stops. \\

\noindent
  0  0  0  0  0  0  0  9  0  0  0  0  0  0  0\\
  0  0  0  0  0  0  1  7  1  0  0  0  0  0  0\\
  0  0  0  0  0  0  2  5  2  0  0  0  0  0  0\\
  0  0  0  0  0  1  1  6  1  0  0  0  0  0  0\\
  0  0  0  0  0  1  2  4  2  0  0  0  0  0  0\\
  0  0  0  0  0  2  1  5  1  0  0  0  0  0  0\\
  0  0  0  0  1  0  3  3  2  0  0  0  0  0  0\\
  0  0  0  0  1  1  2  4  1  0  0  0  0  0  0\\
  0  0  0  0  1  2  0  3  3  0  0  0  0  0  0\\
  0  0  0  0  2  0  2  3  2  0  0  0  0  0  0\\
  0  0  0  1  0  2  1  4  1  0  0  0  0  0  0\\
  0  0  0  1  1  0  2  2  3  0  0  0  0  0  0\\
  0  0  0  1  1  1  1  3  2  0  0  0  0  0  0\\
  0  0  0  1  1  1  2  3  1  0  0  0  0  0  0\\
  0  0  0  1  1  2  1  2  2  0  0  0  0  0  0\\
  0  0  0  1  2  0  3  2  1  0  0  0  0  0  0\\
  0  0  0  2  0  2  2  1  2  0  0  0  0  0  0\\
  0  0  1  0  2  1  1  4  0  0  0  0  0  0  0\\
  0  0  1  1  0  2  2  2  1  0  0  0  0  0  0\\
  0  0  1  1  1  1  2  1  2  0  0  0  0  0  0\\
  0  0  1  1  1  2  0  3  0  1  0  0  0  0  0\\
  0  0  1  1  2  0  2  1  1  1  0  0  0  0  0\\
  0  0  1  2  0  2  0  2  1  1  0  0  0  0  0\\
  0  0  2  0  2  0  2  0  2  1  0  0  0  0  0\\
  0  1  0  2  0  2  0  3  0  1  0  0  0  0  0\\
  0  1  1  0  2  0  2  1  1  1  0  0  0  0  0\\
  0  1  1  1  0  2  0  2  1  1  0  0  0  0  0\\
  0  1  1  1  1  0  2  0  2  1  0  0  0  0  0\\
  0  1  1  1  1  1  0  2  0  2  0  0  0  0  0\\
  0  1  1  1  1  1  1  0  3  0  0  0  0  0  0\\
  0  1  1  1  1  1  1  2  1  0  0  0  0  0  0\\
  0  1  1  1  1  1  2  0  2  0  0  0  0  0  0\\
  0  1  1  1  1  2  0  2  0  1  0  0  0  0  0\\
  0  1  1  1  2  0  2  0  1  1  0  0  0  0  0\\
  0  1  1  2  0  2  0  1  1  1  0  0  0  0  0\\
  0  1  2  0  2  0  1  1  1  1  0  0  0  0  0\\
  0  2  0  2  0  1  1  1  1  1  0  0  0  0  0\\
  1  0  2  0  1  1  1  1  1  1  0  0  0  0  0\\
  1  1  0  1  1  1  1  1  1  1  0  0  0  0  0\\


\section{Discussion}

Formally, the state of the system is described by the status distribution $v(A,t)$. The state at given time step is determined by the parameters $\gamma,p$ and the state at the preceding time step; clearly, the system is Markovian. It is tempting to say that the evolution is local; however, even for given $p$ and $\gamma$ it cannot be reduced to a cellular automaton with neighbourhood limited to nearest neighbors. This can be demonstrated, for example, by an inspection of Table 2, where an outcome of the state 202 happens to be either 2 or 3. \\

Similarly to the computational model \cite{my}, {\it (i)} the mean field approach reproduces the phase transition between the phase where most emotions are positve (small $p$) and the phase where most of them are negative (larger $p$); {\it (ii)} the model reproduces the differentiation in status. However, on the contrary to the model \cite{my}, {\it (iii)} the transition is sharp; {\it (iv)} in the time evolution of the status distribution $v(A)$, a limitation appears which contributes to the polarization of status. When translated to social phenomena, the limitation from above can be interpreted as a kind of 'glass ceiling' \cite{chov}, produced by the group itself. Any claim to eternal happiness limited to believers of one's own religion \cite{bib} can serve as an example of the limitation from below. \\

As stated in the abstract, the mean field approximation applied here decouples the consequence of praising or being praised. In statistical physics such an approximation usually deteriorates the accuracy of results. However, when modeling a social phenomenon we can ask if the decoupling captures some aspects of reality, which is omitted by the direct computational approach. It is straightforward to imagine that the averaged work functions $f_1$ and $f_2$ represent some kind of 'social field', exogenous in the sense 'not reducible to direct face-to-face contacts'. Actually the concept of social field, well established in sociology, has some scientific roots: 'field theory stems from the physical sciences' \cite{ljm}. What does not result directly from face-to-face interpersonal contacts in our model is just the time variation of the mean status $<A>$, as this variation is absent in the purely computational approach \cite{my}, where these contacts are taken into account. \\

\end{document}